\documentclass[11pt,a4paper,twoside]{article}

\usepackage[utf8]{inputenc}	
\usepackage{amssymb}
\usepackage{amsmath}
\usepackage{bm}
\usepackage{graphicx}
\usepackage{subcaption}
\usepackage[title]{appendix}

\usepackage[affil-it,auth-lg]{authblk}
\usepackage{tabularx}
\usepackage{booktabs}
\usepackage{diagbox}
\usepackage[mathscr]{euscript}

\usepackage[dvips]{color} 

\let\OLDthebibliography\thebibliography
\renewcommand\thebibliography[1]{
	\OLDthebibliography{#1}
	\setlength{\parskip}{0pt}
	\setlength{\itemsep}{0pt plus 0.3ex}
}

\allowdisplaybreaks

\begin{document}
	
	\title{The Cosmological Black Hole}
	
\author[1,2]{Zacharias Roupas}
\affil[1]{The British University in Egypt, Faculty of Energy and Environmental Engineering, Sherouk City 11837, Cairo, Egypt}
\affil[2]{Department of Physics, University of Crete, 70013, Herakleion, Greece} 

\date{\vspace{-5ex}}


\maketitle


\begin{abstract}
		We briefly review the recent novel solution of General Relativity, we call the cosmological black hole, firstly discovered in [Roupas, Z. Eur. Phys. J. C 82, 255 (2022)]. A dark energy universe and a Schwartzschild black hole are matched on a common dual event horizon which is finitely thick due to quantum indeterminacy. The system gets stabilized by a finite tangential pressure applied on the dual horizon. The fluid entropy of the system at a Tolman temperature identified with the cosmological horizon temperature is calculated to be equal with the Bekenstein-Hawking entropy.
	\end{abstract}

	A universe hosted inside a black hole within a general relativistic framework is an old idea  \cite{1972Natur.240..298P,1981NCimL......161G,1989PhLB..216..272F}. Related, but different, is the idea that a black hole singularity may get regularized, generating black hole mimickers like gravastars or dark energy stars (e.g. see \cite{Ansoldi:2008jw} and references therein).

	Poisson and Israel \cite{1988CQGra...5L.201P} have pointed out that when an interior de Sitter core is matched with an exterior Schwarzschild spacetime exactly at the radius $r_{\rm H}$ where the cosmological event horizon and the black hole event horizon coincide then a tangential pressure is applied to the horizon equal to $P_{\rm T} = \frac{1}{2} \rho_0 \delta \left(\frac{r}{r_{\rm H}} - 1 \right)$, where $\rho_0$ denotes the constant dark energy (de Sitter) mass-energy density. They further observed that this tangential pressure diverges for an observer at any proper distance $\Delta s$
	\begin{equation}\label{eq:P_T_divergence}
		P_{\rm T} = \frac{1}{2}\rho_0 \delta\left(g_{rr}^{-1/2} \frac{\Delta s}{r_{\rm H}} \right) = \frac{1}{2}\rho_0 r_{\rm H}  g_{rr}^{1/2} \delta(\Delta s)
		\rightarrow \infty, \;\text{for} \; r = r_{\rm H}.
	\end{equation}
	
	Thus, the Schwarzschild singularity of curvature in the center is replaced by a pressure singularity on the horizon if an exterior Schwarzschild black hole spacetime is matched with an interior de Sitter spacetime exactly on a common event horizon
	\begin{equation}\label{eq:r_H}
		r_{\rm H} \equiv \frac{2GM_\bullet}{c^2} = \sqrt{\frac{3c^2}{8\pi G \rho_0}} ,
	\end{equation}
where $M_\bullet$ denotes the total mass-energy of the black hole 
	\begin{equation}
M_\bullet = \frac{4}{3}\pi  r_{\rm H}^3 \rho_0.
\end{equation}

However, we have recently remarked \cite{2022EPJC...82..255R} that according to quantum physics the boundary $r_{\rm H}$ of a finitely distributed $M_\bullet$ cannot be specified with accuracy greater than the Compton wavelength  
\begin{equation}
	\Delta r_{\rm H} \gtrsim \frac{h}{M_\bullet c}.
\end{equation}
Therefore there should exist some length-scale $\alpha$ which specifies the quantum fuzziness of the horizon 
$
	\Delta r_{\rm H}=\alpha 
$
in this case. We shall not be concerned with the exact value of $\alpha$ (it may attain any value between the Compton wavelength and the Planck scale), but consider that for an astrophysical black hole it is
\begin{equation}\label{eq:varepsilon}
	\varepsilon \equiv \frac{\alpha}{r_{\rm H}} \ll 1.
\end{equation}
We have discovered solutions of General Relativity \cite{2022EPJC...82..255R} that regularize the Poisson-Israel solution, namely we regularize the tangential pressure on the horizon which we assume now to be fuzzy with a finite width 
\begin{equation}
	r_{\rm H}-\frac{\alpha}{2} \leq r \leq r_{\rm H}+\frac{\alpha}{2}.
\end{equation}

	Assuming the static, spherically symmetric ansatz 
	\begin{equation}\label{eq:ds2_ansatz}
		ds^2 = - h(r) c^2 dt^2 + h(r)^{-1} dr^2 + r^2 d\Omega,
	\end{equation}
	and an anisotropic energy momentum tensor $T_0^{~0} = -\rho(r) c^2$, $T_1^{~1} = P_{\rm r} (r)$, $T_2^{~2}=T_3^{~3} = P_{\rm T}(r)$, the Einstein equations 
	\begin{equation}\label{eq:einstein}
		R_\mu^{~\nu} - \frac{1}{2} R_\sigma^{~\sigma} \delta_\mu^{~\nu} = \frac{8\pi G}{c^4}T_\mu^{~\nu}
	\end{equation}
	give 
	\begin{equation}
		\label{eq:dm_methods}
		h(r) = 1 - \frac{2 G m(r)}{r c^2}, \quad
		\frac{dm(r)}{dr} = 4\pi r^2 \rho(r), \quad 
		P_{\rm r}(r) = - \rho(r) c^2, \quad
		P_{\rm T}(r) = - \rho(r) c^2 - \frac{1}{2} r .
	\end{equation}
We have discovered the following analytical solutions
\begin{equation} 
	\label{eq:solutions}
	\rho(r) = \left\lbrace
	\begin{array}{ll}
		\rho_0 &, \; r \leq r_{\rm H} - \frac{\alpha}{2} \\
		\rho_{(-)}(r) &, \;  r_{\rm H} - \frac{\alpha}{2} \leq r \leq r_{\rm H} \\
		\rho_{(+)}(r) &, \;  r_{\rm H} \leq r \leq r_{\rm H} + \frac{\alpha}{2}  \\
		0 &, \; r \geq r_{\rm H} + \frac{\alpha}{2} \\
	\end{array}
	\right. 
	,
	\quad	
	\rho_{(\pm)}(r) = \rho_0\sum_{n=0}^N A^{(\pm)}_n (\varepsilon) \, x(r)^n ,\;
	x \equiv \frac{r - r_{\rm H}}{\alpha} \in {[-\frac{1}{2},+\frac{1}{2}]}.
\end{equation}
Proper choices of $A^{(\pm)}_n(\varepsilon)$ ensure that the density and consequently the metric are continuous and have continuous derivatives. The maximum order of the continuous derivatives can be arbitrarily high.
We ensure this by demanding to hold the following conditions
\begin{align}
	\label{eq:cond_1}	
	&\rho_{(-)}(r_{\rm H}-\frac{\alpha}{2}) = 	\rho_0 , 	\quad
	\rho_{(-)}(r_{\rm H}) = \rho_{(+)}(r_{\rm H}) , 
	\quad 
	\rho_{(+)}(r_{\rm H}+\frac{\alpha}{2}) = 	0 , 	\\
	\label{eq:cond_2}	
	&\frac{d^{(k)}\rho_{(-)} (r_{\rm H}-\frac{\alpha}{2})}{dr^k} = 	0 ,
	\;
	\frac{d^{(k)}\rho_{(-)} (r_{\rm H})}{dr^k} = \frac{d^{(k)}\rho_{(+)} (r_{\rm H})}{dr^k}, 	
	\; 
	\frac{d^{(k)}\rho_{(+)} (r_{\rm H}+\frac{\alpha}{2})}{dr^k}  = 0 , 
	\quad 
	k = 1, 2, \ldots, K
	\\
	\label{eq:cond_3}	
	&\int_{r_{\rm H}-\frac{\alpha}{2}}^{r_{\rm H}} 4\pi \rho_{(-)}r^2 dr + \int_{r_{\rm H}}^{r_{\rm H}+\frac{\alpha}{2}} 4\pi \rho_{(+)}r^2 dr 	= \frac{4}{3}\pi \rho_0\left( r_{\rm H}^3 - \left( r_{\rm H} - \frac{\alpha}{2} \right)^3 \right) .
\end{align}
We impose $K\geq 1$ to ensure that the metric along with its first and second derivatives are continuous, satisfying Lichnerowicz-Darmois junction conditions (which require continuity of the metric up to the first derivative). The condition (\ref{eq:cond_3}) ensures that condition (\ref{eq:r_H}) holds, namely that the cosmological and black hole event horizon coincide and thus $M_\bullet=\frac{4}{3}\pi r_{\rm H}^3$. We also demand that the weak energy condition holds $\rho^\prime \leq 0$, $\rho\geq 0$, $P_r + \rho \geq 0$, $P_T+\rho\geq 0$.

\begin{figure}[!tb]
	\begin{center}
		\begin{subfigure}{0.45\textwidth}
			\includegraphics[scale = 0.5]{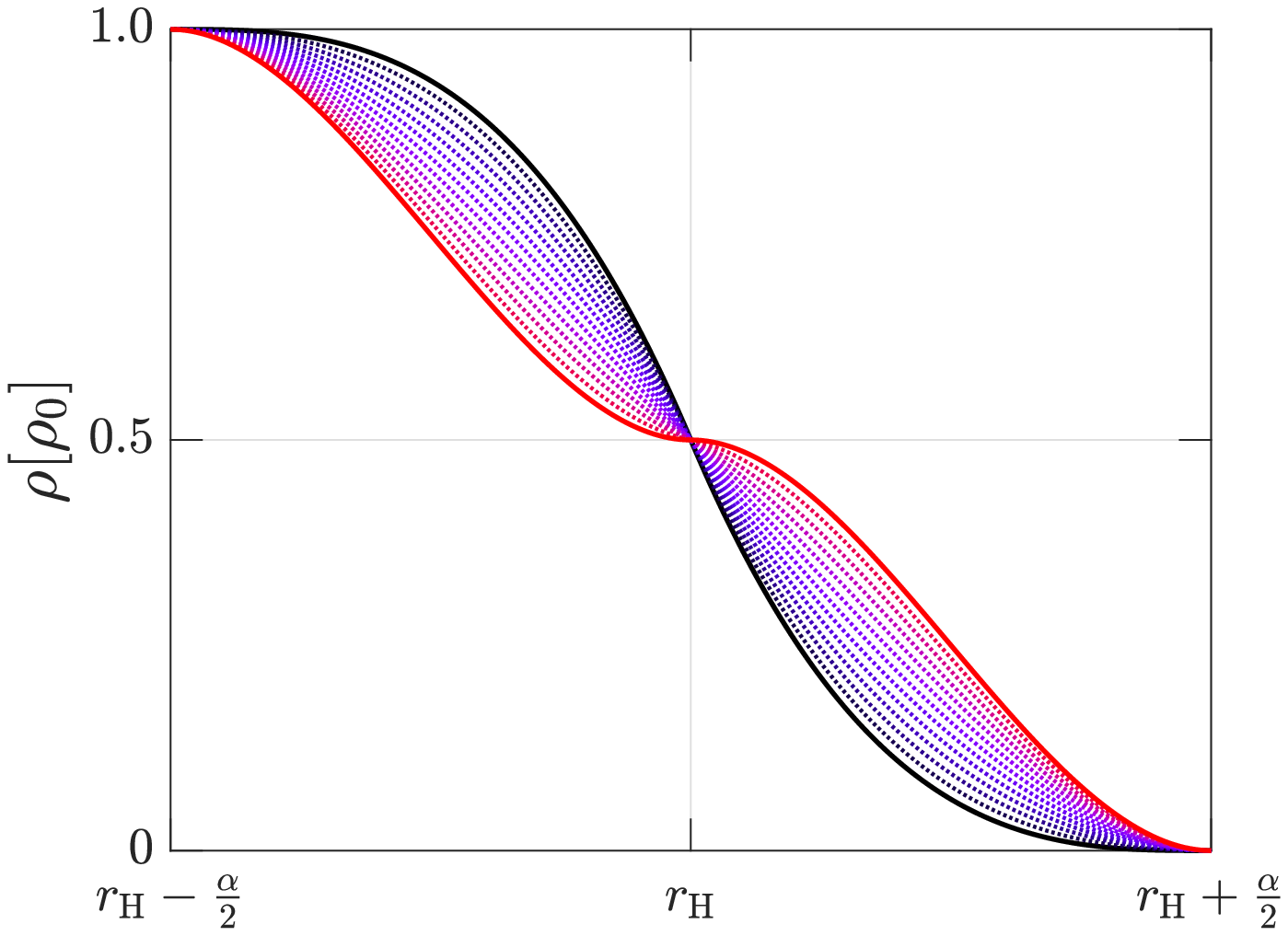}  
			\caption{Density.}
			\label{fig:rho_1st_order_spectrum}
		\end{subfigure}
		\begin{subfigure}{0.45\textwidth}
			\includegraphics[scale = 0.5]{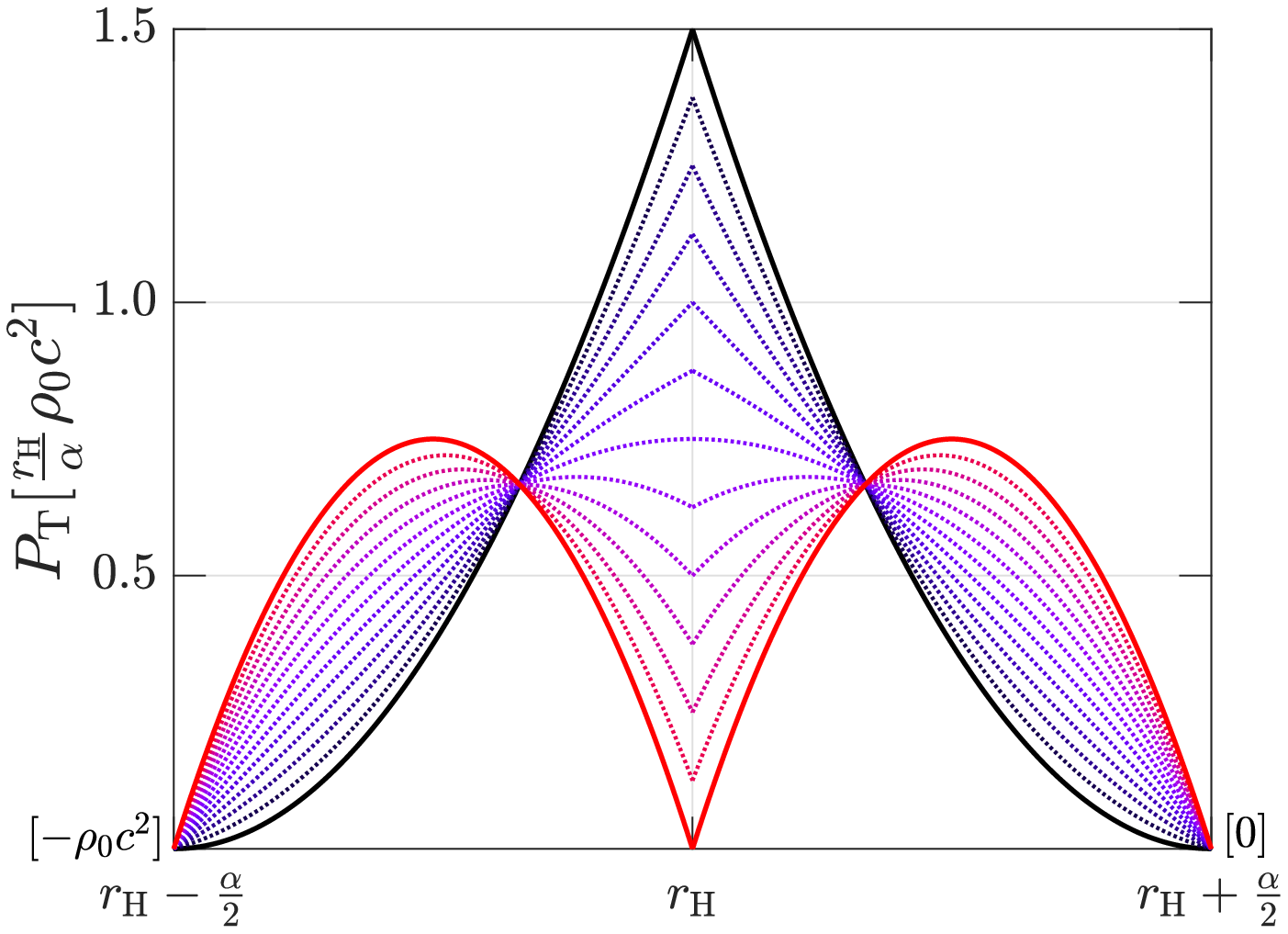}  
			\caption{Tangent pressure.}
			\label{fig:P_T_1st_order_spectrum}
		\end{subfigure}
	\captionsetup{justification=raggedright}
		\caption{The cosmological black hole spectrum,  $K=1$, $N=3$, of density and tangential pressure inside the horizon. The values of $P_{\rm T}$ on the boundaries are mentioned separately in brackets. Each curve represents a solution of the Einstein equations with the same energy and entropy. }
		\label{fig:spectrum}
	\end{center}
\end{figure}

A spectrum of solutions for $K=1$, $N=3$ is 
\begin{equation}
\begin{array}{l}
	A_0^{(-)} = A_0^{(+)} = \frac{ 60 - (18  + B )\varepsilon + 2 \varepsilon^2}{4 (\varepsilon^2 - 4 \varepsilon + 30)} = \frac{1}{2} + \mathscr{O}(\varepsilon)
	\\
	A_1^{(-)} = A_1^{(+)} = \frac{-240 + 30 B + 72\varepsilon + (B - 8)\varepsilon^2 }{4(\varepsilon^2 - 4\varepsilon + 30)} = \frac{B-8}{4} + \mathscr{O}(\varepsilon)
	\\
	A_2^{(-)} = \frac{-60 + 30B + (78+ 3B)\varepsilon + (B - 2)\varepsilon^2 }{\varepsilon^2 - 4\varepsilon + 30} = B-2 + \mathscr{O}(\varepsilon)
	\\
	A_2^{(+)} = -\frac{-60 + 30B + (18 - 3B)\varepsilon + (B - 2)\varepsilon^2}{\varepsilon^2 - 4\varepsilon + 30} = -(B-2) + \mathscr{O}(\varepsilon)
	\\
	A_3^{(-)} =  \frac{30B + (80 + 4B)\varepsilon + B\varepsilon^2}{\varepsilon^2 - 4\varepsilon + 30} = B + \mathscr{O}(\varepsilon)
	\\
	A_3^{(+)} =  B 	
\end{array} 
,\quad \text{with}\; -4 \leq B \leq 8 .
\end{equation}
Additional analytical solutions may be found in the Appendix of \cite{2022EPJC...82..255R} along with additional plots. In Figure \ref{fig:spectrum} we depict the solution $K=1$, $N=3$. There can always be found solutions as long as $N \geq (3K+2)/2$. We have calculated analytical solutions up to $K=15$, that means continuity up to the the sixteenth derivative of the metric. 

Let us now calculate the fluid entropy of the cosmological black hole. The work of the gravitational force under a spherical deformation (no pure shear) is 
\begin{equation}
	dW = P dV, \quad {\rm where}\; P = \frac{1}{3}(P_{\rm r} + P_{\rm T} + P_{\rm T}) = P_{\rm r} + \frac{2}{3}(P_{\rm T} - P_{\rm r}).
\end{equation}
There is an additional, to $P_{\rm r}$, contribution $\frac{2}{3}(P_{\rm T} - P_{\rm r})$ coming from the tangential pressure due to the stretching forces on the surface of a fluid sphere during any spherical deformation. Therefore the pressure $P$ and not only $P_r$ should contribute to the entropy density.
The thermodynamic Euler relation for zero chemical potential is written thus as
\begin{equation}\label{eq:Euler}
	Ts = \rho c^2 + P \Rightarrow s = -\frac{c^2}{3T} r\rho^\prime,
\end{equation} 
where $s=s(r)$ is the total entropy density, including the tangential contribution, and we used equation (\ref{eq:dm_methods}). We denote $T=T(r)$ the local temperature, which obeys the Tolman law
\begin{equation}\label{eq:Tolman}
	T(r)\sqrt{g_{tt}(r)} = T_0 = {\rm const.}
\end{equation}
where the constant $T_0$ is called the Tolman temperature and is the thermodynamic quantity conjugate to the energy in General Relativity. It is the temperature of the system, at any point, measured by an observer at infinity.
The interior does not contribute to the fluid entropy $Ts_{\rm interior} = \rho_0 c^2 + P_{\rm interior} = \rho_0 c^2 - \rho_0 c^2 = 0$.
Thus, the total fluid entropy is 
\begin{equation}
	S = \int_{r_{\rm H}-\frac{\alpha}{2}}^{r_{\rm H} + \frac{\alpha}{2}} s(r) \sqrt{g_{rr}}4\pi r^2dr = 
	- \frac{c^2}{3T_0}\int_{r_{\rm H}-\frac{\alpha}{2}}^{r_{\rm H} + \frac{\alpha}{2}} \rho^\prime 4\pi r^3dr = \frac{4\pi c^2}{3T_0}\rho_0 \left(r_{\rm H}-\frac{\alpha}{2} \right)^3 
	+ \frac{ c^2}{T_0} \int_{r_{\rm H}-\frac{\alpha}{2}}^{r_{\rm H} + \frac{\alpha}{2}} \rho \,4\pi r^2dr  = \frac{4\pi c^2}{3}\frac{\rho_0 r_{\rm H}^3}{T_0}
\end{equation}
for all solutions (\ref{eq:solutions}) subject to the conditions (\ref{eq:cond_1})-(\ref{eq:cond_3}). We finally get
\begin{equation}\label{eq:entropy}
	S =  \frac{M_\bullet c^2}{T_0},
\end{equation}
where we used equation (\ref{eq:r_H}) that identifies the coincidence of cosmological and black hole event horizons.
Note that this result holds for any choice of $\alpha$ and for all solutions of the cosmological black hole spectrum (\ref{eq:solutions}). All such solutions with the same total mass-energy, correspond also to the same entropy. 

If $\alpha$ is the quantum indeterminacy of the event horizon, the Tolman temperature $T_0$ is equal to the cosmological temperature $T_{\rm dS}$ 
\begin{equation}\label{eq:T_0}
	T_0 = T_{\rm dS} \equiv \frac{\hbar c}{2\pi r_{\rm H}} =  \frac{\hbar c^3}{4\pi G M_\bullet} \equiv 2T_{\rm BH}, 
\end{equation}
By direct substitution of the cosmological temperature in entropy (\ref{eq:entropy})
\begin{equation}\label{eq:S}
	S = \frac{4\pi G}{\hbar c}M_\bullet^2 = \frac{4\pi r_{\rm H}^2}{4\hbar G/c^3} \equiv S_{\rm BH},
\end{equation}
we reach the intriguing conclusion that the fluid entropy equals the Bekenstein-Hawking entropy $S_{\rm BH}$. 

\begin{figure}[!tb]
	\begin{center}
		\begin{subfigure}{0.45\textwidth}
			\includegraphics[scale = 0.5]{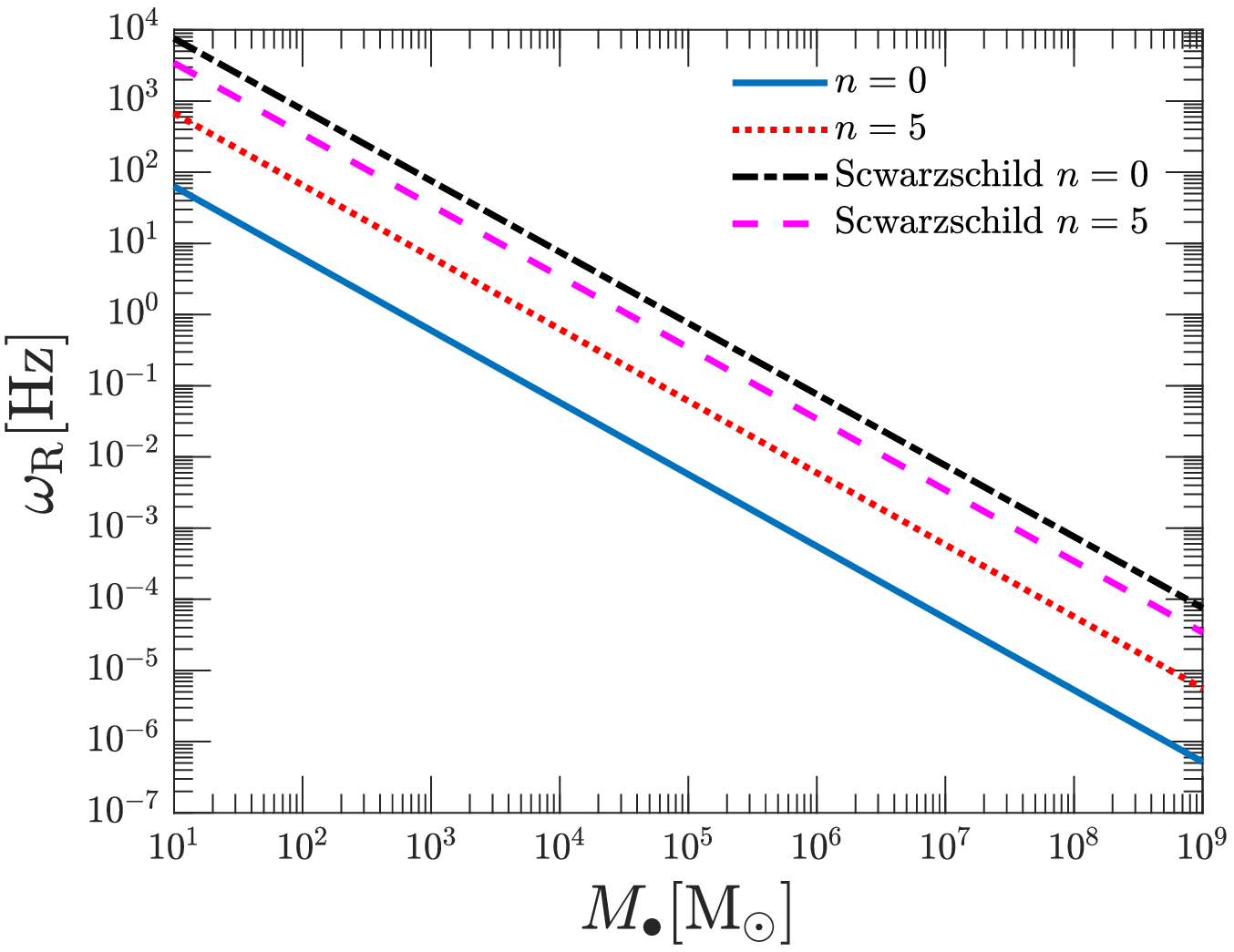}  
			\label{fig:omegaR_M_Hz_l=2}
		\end{subfigure}
		\begin{subfigure}{0.45\textwidth}
			\includegraphics[scale = 0.5]{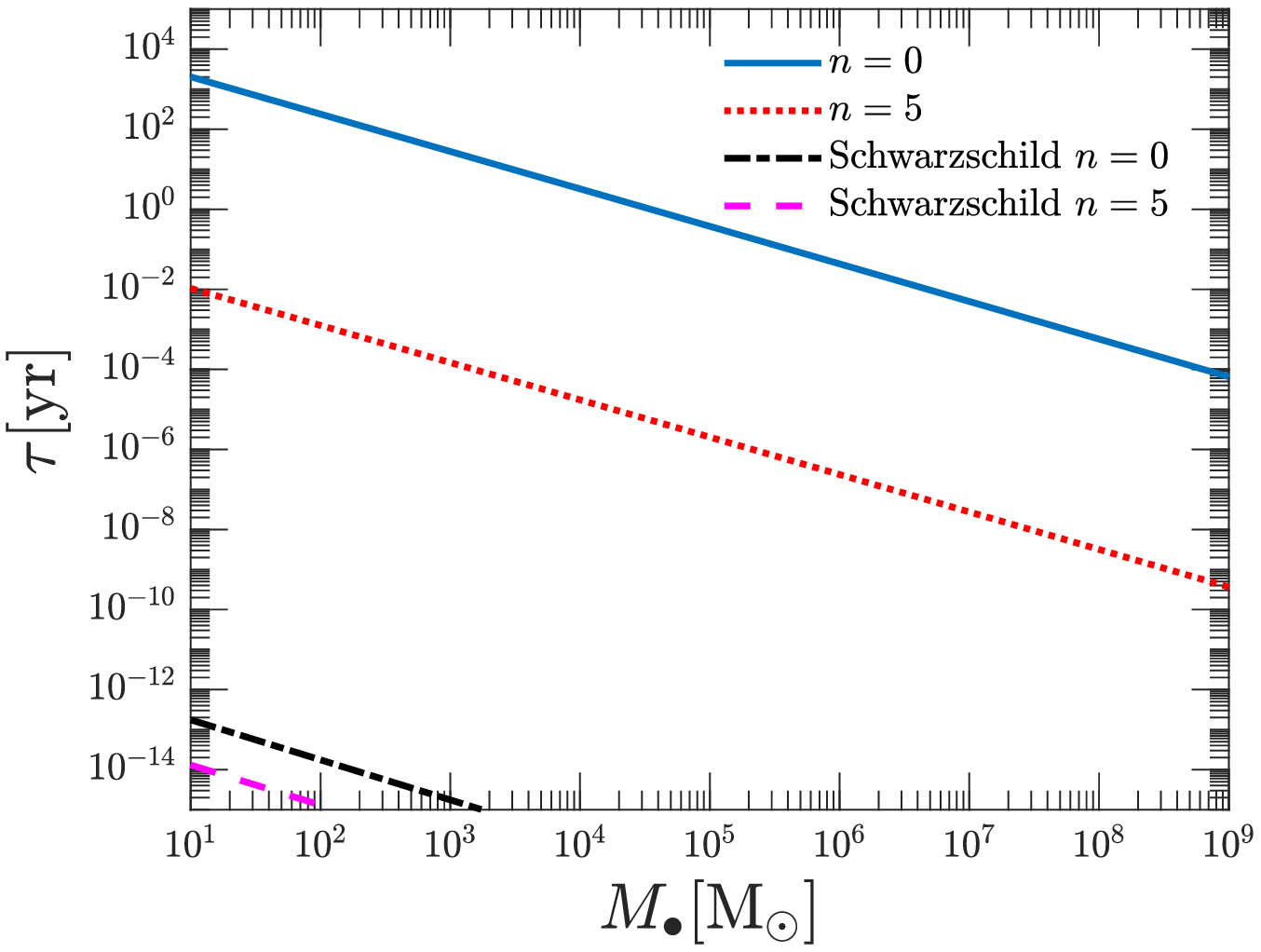}  
			\label{fig:tau_M_yr_l=2}
		\end{subfigure}
	\captionsetup{justification=raggedright}
		\caption{The frequency (left panel) and damping period (right panel) of the quasinormal modes $n=0$, $n=5$  for $\ell=2$ of the cosmological black hole with respect to its mass $M_\bullet$. These apply to all solutions while differences in the precise value of $\varepsilon$ are also negligible. There are additionally depicted the values of Schwarzschild black holes for comparison.}
		\label{fig:omega-tau_l=2}
	\end{center}
\end{figure}

Finally, in Ref \cite{2022EPJC...82..255R} we have shown that radial perturbations cannot induce instabilities and  we have calculated the quasinormal modes of the cosmological black hole. These are depicted in Figure \ref{fig:omega-tau_l=2}. 

We see that the quasi-normal modes of cosmological black holes are distinctively different than the ones of Schwarzschild black holes. Still, LIGO-Virgo cannot disciminate between cosmological and Schwarzschild black holes, because of the well-known mode camouflage mechanism \cite{PhysRevD.79.064016,2014PhRvD..90d4069C}. This refers to the fact that the ringdown waveform of a binary black hole merging is dominated initially by spacetime fluctuations in the region of the external null geodesic, which is common in regular and singular black holes. Additionally, we see that the mode frequencies of astrophysical cosmological black holes, namely $10^{-6}{\rm Hz} \lesssim \omega_{\rm R} \lesssim 10{\rm Hz}$, lie outside the frequency's range detectability of LIGO-Virgo. Therefore it is not excluded the possibility that the LIGO-Virgo detections are cosmological black holes. 

It is furthermore evident from Figure \ref{fig:omega-tau_l=2}  that for $M_\bullet \gtrsim 10^4{\rm M}_\odot$ the fundamental mode of cosmological black hole fluctuations lies within the frequency detectability range ($\sim 10^{-1}-10^{-5}{\rm Hz}$) of the LISA space interferometer. Since the lifetime of the fundamental mode of a cosmological black hole is sufficiently larger than the Schwarzschild one (right panel of Figure \ref{fig:omega-tau_l=2}), it is possible that LISA can detect the echo of the cosmological black hole fluctuations provided LISA is sensitive enough. This is intriguing subject for future research.

\bibliography{2022_CosmoBH_arxiv}

\end{document}